\documentstyle[12pt]{article}
\begin{document}
\input epsf.tex
\def\a{\alpha}
\def\b{\beta}
\def\ch{\chi}
\def\d{\delta}
\def\e{\epsilon}
\def\E{{\cal E}}
\def\f{\phi}
\def\g{\gamma}
\def\h{\eta}
\def\i{\iota}
\def\j{\psi}
\def\k{\kappa}
\def\l{\lambda}
\def\m{\mu}
\def\n{\nu}
\def\o{\omega}
\def\p{\pi}
\def\q{\theta}
\def\r{\rho}
\def\s{\sigma}
\def\t{\tau}
\def\u{\upsilon}
\def\x{\xi}
\def\z{\zeta}
\def\D{\Delta}
\def\F{\Phi}
\def\G{\Gamma}
\def\J{\Psi}
\def\L{\Lambda}
\def\O{\Omega}
\def\P{\Pi}
\def\S{\Sigma}
\def\U{\Upsilon}
\def\X{\Xi}
\def\T{\Theta}
\def\vf{\varphi}
\def\ve{\varepsilon}
\def\cC{{\cal X}}
\def\cD{{\cal Y}}
\def\Ab{\bar{A}}
\def\gi{g^{-1}}
\def\li{{ 1 \over \l } }
\def\lb{\l^{*}}
\def\zb{\bar{z}}
\def\ub{u^{*}}
\def\vb{v^{*}}
\def\Tb{\bar{T}}
\def\sech {{\rm{sech}}}
\def\csch {{\rm{csch}}}
\def\cn{{\rm{cn}}}
\def\dn{{\rm{dn}}}
\def\sn{{\rm{sn}}}
\def\be{\begin{equation}}
\def\ee{\end{equation}}
\def\ben{\begin{eqnarray}}
\def\een{\end{eqnarray}}
\def\lt{\tilde{\lambda}}
\def\ben{\begin{eqnarray}}
\def\een{\end{eqnarray}}
\def\pb{\partial_{\zb}}
\def\pp{\partial_{z}}
\def\s{\sigma}
\def\cw{\psi^{cw}}
\def\Et{\tilde{E}}
\def\Im{{\rm{Im}}}
\def\Re{{\rm{Re}}}
\hsize=15truecm
\addtolength{\topmargin}{-0.6in}
\vsize=26.5truecm
\hoffset=-0.3in

\rightline{\today} 
\vskip2cm
\centerline{\Large\bf Soliton on a Cnoidal Wave Background }
\centerline{\Large\bf in the coupled nonlinear Schr\"{o}dinger equation }
\vskip 1cm
\centerline{
H. J. Shin\footnote{ Electronic address; hjshin@khu.ac.kr }}
\vskip 3mm
\centerline{Department of Physics and Research Institute of Basic
Science}
\centerline{Kyung Hee University, Seoul 130-701,  Korea}
\vskip 2cm
\centerline{\bf ABSTRACT}
\vskip 5mm

An application of the Darboux transformation on a cnoidal wave background
in the coupled nonlinear Schr\"{o}dinger equation
gives a new solution which describes a soliton moving on a cnoidal wave. 
This is a generalized version of the
previously known soliton solutions of dark-bright pair. Here a dark soliton resides on a 
cnoidal wave instead of on a constant background. It also
exhibits a new types of soliton solution
in a self-focusing medium, which describes a breakup of a generalized dark-bright pair into another generalized dark-bright pair
and an ``oscillating" soliton.
We calculate the shift of the crest of the cnoidal wave along a soliton and the moving direction 
of the soliton on a cnoidal wave.

\vskip 5mm

\newpage

\setcounter{footnote}{0}

\vskip 5mm
\section{Introduction}
In this paper, we consider a system of coupled nonlinear Schr\"{o}dinger (CNLS) equations  \cite{man,ag,zak}
\be
\pb \psi_{k} = -i \pp^2 \psi_{k} -2i \s  (|\psi_{1}|^2+|\psi_{2}|^2
)\psi_{k} ~ ; ~ k=1,2.
\label{vnls}
\ee
These equations, which were
named as the Manakov model, are important for a number of physical applications like multi-frequency
and/or two different polarizations of light in a fiber  \cite{ber,shin,kivs}. 
Nowadays, there exist vast amount of exact solutions of this system including exact vector solitons \cite{man,shin}, 
bound solitary waves \cite{hioe}, and periodic solutions \cite{hioe,flor,rom,shin11}.  
Especially, quasi-periodic solutions
in terms of N-phase theta functions for the Manakov model are derived in \cite{adams},
while a series of special solutions are given in \cite{alp,poly,poru,pul}. 

Recently, this equation attracts new attention as
it can describe the localized states in optically induced refractive index gratings \cite{efr,fle,des}.
Strong incoherent interaction of such a grating with a probe beam facilitates the formation of a noble
type of a composite optical soliton, where one of the components (described by $\j_1$ in Eq. (\ref{vnls}))
creates periodic photonic structure, while the other component ($\j_2$) experiences
Bragg reflection from this structure and forms gap solitons. Optically induced lattices are very exciting development
which open up creating dynamically reconfigurable photonic structures. It was shown that the CNLS
equation describes the propagation of two incoherently interacting beams in a photorefractive
crystal in the limit of weak saturation regime \cite{led,des}.

To analyze the behavior of the soliton in the optically induced lattices, we need a
solution of the CNLS equation that describes a soliton moving on a cnoidal wave.
The integrability of the CNLS equation shown by Manakov provides us to obtain various type of solutions.
Especially, one can obtain the bright soliton
in a focusing medium  by applying the inverse scattering method (ISM) \cite{Manakov}. On the other hand, in the
case of nonvanishing background fields, the inverse scattering method is technically
highly involved and only the dark solitons of the simplest one-component case have been found in this way.
In the two component case, the Hirota method has been adopted to obtain dark solitons instead of the ISM 
\cite{Radhak,Sheppard}.
The Hirota method, however,
does not provide a way to construct a soliton solution moving on a cnoidal wave background.
Interestingly, the (soliton + cnoidal wave) solution can be obtained from the general quasi-periodic solutions
of N-phase theta functions,
by taking degenerate limit of the 2-phase solution. In fact, Ref. \cite{ab} applied this procedure
to obtain a solution of the single-component nonlinear Schr\"{o}dinger equation (NLSE). 
But solutions from the N-phase theta functions have
the so-called ``effectivization" problem, which is related to extracting the physical solutions by
taking proper initial conditions \cite{kam,shin11}. Much more, these solutions have 
rather complicated form, which makes them difficult to apply to real situations. 

In this paper, we employ a simple, but very powerful soliton
finding technique based on the Darboux transformation (DT). This method is
essentially equivalent to the ISM, but avoids mathematical
technicalities of the ISM. Section 2 introduces the main method including the DT and Sym's solution.
Section 3 analyzes the characteristics of obtained solutions in the case of focusing medium.
Section 4 is devoted to the case of defocusing medium. Section 5 contains a short discussion and
Appendices give some proofs of equations in the main text.

\section{The Method}
\subsection{Lax Pair}

We first bring the CNLS equation into a matrix form in
terms of $3 \times 3 $ matrices $E, T$ and $\tilde{E} = [T, ~ E]$,
\be
E=\pmatrix{0 & \psi_{1} &  \psi_{2} \cr
           -\s \psi_{1}^{*} & 0 & 0 \cr
           -\s \psi_{2}^{*} & 0 & 0 },
~~~
T = \pmatrix{i/2 & 0 & 0 \cr
           0 & -i/2 &  0 \cr
           0 & 0 & -i/2 },
\label{ET}
\ee
such that
\be
\pb E = -\pp^2 \tilde E +2E^{2} \tilde E.
\label{nlseq}
\ee
One can readily check that the components of Eq. (\ref{nlseq})
are indeed equivalent to the Manakov equation in Eq. (\ref{vnls}).
The signature $\s $ is either $1$ or $-1$ depending on whether the
group velocity dispersion is abnormal ($\s =1 $) or normal
($\s=-1 $), or the waveguide is self-focusing ($\s =1 $) or
self-defocusing($\s = -1)$. One advantage of using matrices is
that we can write down the associated linear equation (Lax pair);
\be
(\pp + E +\l T ) \Psi = 0,~~
(\pb + E \tilde E  -\pp \tilde E -\l E -\l^2 T ) \Psi = 0,
\label{nlsle}
\ee
where $\l $ is an arbitrary complex number and
$\Psi (z, \zb, \lambda )$ is a three component vector.

\subsection{Darboux Transformation}
The following cnoidal wave solution of the CNLS equation properly describes
the background periodic wave in optically induced refractive index gratings. For the case of focusing medium ($\s=1$), it is
\be
\j_1 ^{c}(z, \zb)=p~ {\rm dn} (\ch,k) e^{i \z}, ~\j_2 ^c =0,
\label{j121}
\ee
where $\ch=-p(z - v \zb), ~\z=[-v z/2-p^2 (2-k^2 ) \zb+v^2 \zb/4]$.
The defocusing case ($\s=-1$) is obtained by substituting $p \rightarrow i p, k \rightarrow i k$
in Eq. (\ref{j121}) of the focusing case. 
\footnote{When we substitute $p \rightarrow ip$ only, it becomes a {\it singular} solution of the defocusing
CNLS equation.}
Here dn, sn is the standard Jacobi elliptic function.
And $v$ is the velocity of
the cnoidal wave and $k \in (0,1)$ is the modulus of the Jacobi function.
As far as elliptic functions are involved we employ terminology and
notation of Ref. \cite{jms} without further explanations.
To describe the characteristics of the composite optical soliton formed on the background grating,
we need a soliton solution superposed on the cnoidal wave.
To obtain a superposed solution of (soliton + cnoidal wave)  using the Darboux transformation, we need to find
a solution of the linear equations (\ref{nlsle}) with $\j_i,i=1,2$
given by Eq. (\ref{j121}). We denote the solution of the linear
equation as a three component vector, $\Psi=\pmatrix{s_0 \cr s_1 \cr s_2}$.
Then a new solution of (soliton + cnoidal wave)
is constructed using the Darboux transformation \cite{mat,Park1,shin1}, which is
\be
\j_i^{c-s} (z,\zb) = \j_i ^c (z,\zb)+i(\l -\l^* ) {\s s_0 s_i ^* \over |s_0|^2 +\s \sum_{j=1,2} |s_j|^2},~i=1,2 .
\label{newj}
\ee
Using that $s_i$ satisfy the associated linear equations in Eq. (\ref{nlsle}), it can be directly checked 
that $\j_i ^{c-s},~i=1,2$ in Eq. (\ref{newj}) is a new solution of the CNLS equation.

\subsection{Sym's Solution}
Explicitly, the solution of the linear equations in Eq. (\ref{nlsle}) for $\s=1$ can be 
written down as
\ben
s_0 &=& e^{i \z/2} [M e^{i\D}
\q_2 ({-iu \over 2K}) \q_0 ({\ch+iu \over 2K})-N e^{-i \D} 
\q_1 ({-iu \over 2K}) \q_3 ({\ch-iu \over 2K})]/\q_0 ({\ch \over 2K}), \nonumber \\
s_1 &=& e^{-i \z/2} [-M e^{i \D}
\q_1 ({-iu \over 2K}) \q_3 ({\ch+iu \over 2K})+N e^{-i \D} 
\q_2 ({-iu \over 2K}) \q_0 ({\ch-iu \over 2K})]/\q_0 ({\ch \over 2K}), \nonumber \\
s_2 &=& C e^{i \l z/2-i \l^2 \zb /2},
\label{sollin}
\een
where $M, N, C$ are arbitrary complex numbers and  $\D=-\g \zb+ \b \ch$ with
\ben
\g &=& -{p^2 \over 2} [ {\rm dn}^2 (u, k') + {{\rm cn}^2 (u,k') \over {\rm sn}^2 (u,k')}], 
\nonumber \\
\b &=& {d \over du} \ln \q_0 ({iu \over 2K}) +{1 \over 2} {{\rm dn}(u,k') {\rm cn}(u,k')
\over {\rm sn}(u,k')} +{{\rm sn}(u,k') {\rm dn}(u,k') \over {\rm cn}(u,k')},
\label{gb}
\een
and $K$ and $E$ are complete elliptic integrals of the first and the second kinds, respectively.
In the above formula, a complex parameter $u$ is related to the DT parameter $\l$ as following,
\be
\l = {v \over 2} +p {{\rm dn}(u,k') {\rm cn}(u,k')
\over {\rm sn}(u,k')}.
\label{lambda}
\ee
Sym first introduced this solution in the description of vortex motion in hydrodynamics  \cite{sym}.
It was then applied to an NLSE-related problem in Ref. \cite{shinpre}. The proof of Sym's solution is given 
in the Appendix I. 
Finally, solutions of the case $\s=-1$ is obtained by substituting $p \rightarrow i p,~k \rightarrow i k$ in Eq. (\ref{sollin}).

\section{Solutions of Self-focusing Medium}
\subsection{A Soliton crossing the Cnoidal Wave}
Figure 1 shows (dark soliton + cnoidal wave) $|\j_1|$, and a bright soliton $|\j_2|$. 
It is obtained using Eqs. (\ref{j121}), (\ref{newj}),
(\ref{sollin}), (\ref{gb}) and (\ref{lambda}) and taking $N=0$. 
Other parameters for the figure are $v=0, k=0.9, p=1.3, u=-0.38+0.63i, M=1, C=0.3.$
The figure is drawn using MATHEMATICA, which is also used in checking that
the solution indeed satisfies the equation of motion (\ref{vnls}).
It shows the characteristic dark-bright pair
soliton of CNLS equation where the dark soliton resides on a cnoidal background.
This case becomes the well-known dark-bright soliton pair when we take $k=0$ \cite{kivs}.
\begin{figure}
\centerline{\epsfxsize 5. truein \epsfbox {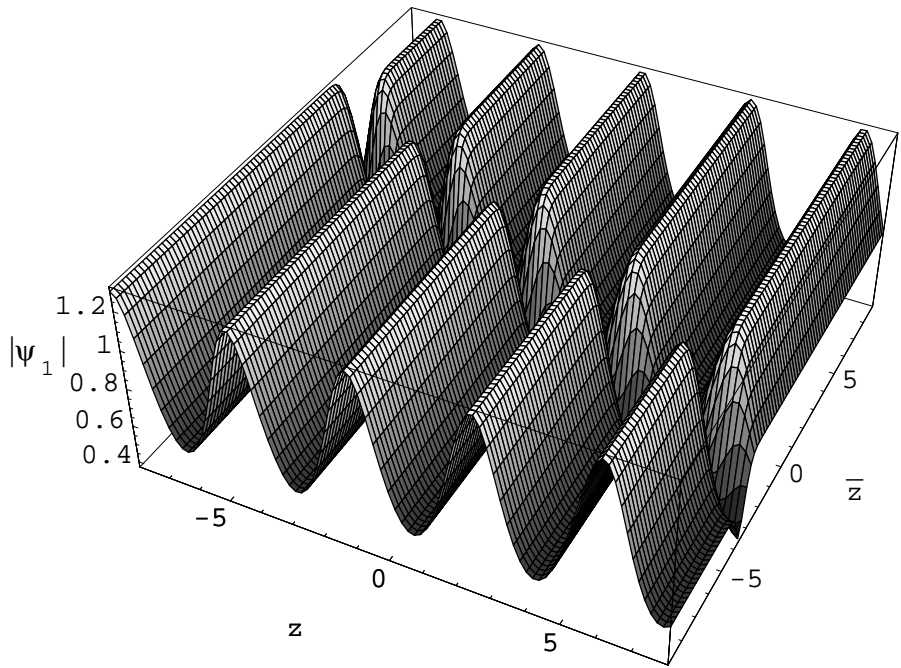}}
\centerline{\epsfxsize 5. truein \epsfbox {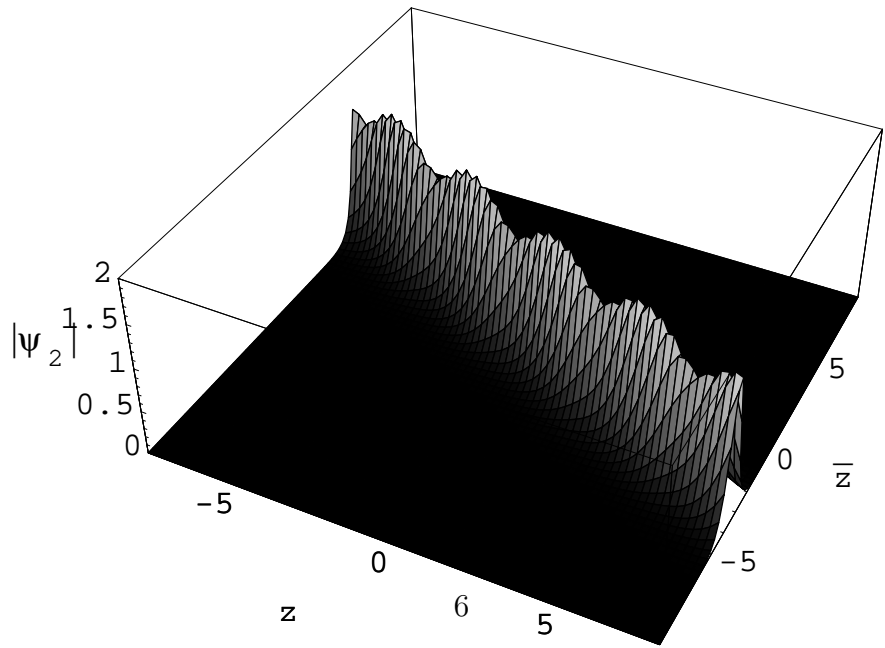}}
\caption{Dark-bright soliton pair of CNLS equation:  $|\psi_1|$ shows a dark soliton residing on a cnoidal background. 
$|\psi_2|$ shows a bright soliton.
The parameters are $v=0, k=0.9, p=1.3, u=-0.38+0.63i, M=1, N=0, C=0.3.$}
\end{figure}

The dark-bright pair soliton moves along a line which satisfies $|s_2| \sim |s_0^{(M)}| \sim |s_1^{(M)}|$ in Eq. (\ref{sollin}).
Here, $|s_0^{(M)}|$ or $|s_1^{(M)}|$ mean $M$-coupled terms in Eq. (\ref{sollin}).
Thus $|s_0^{(M)}|$ is obtained from $|s_0|$ in Eq. (\ref{sollin}) by taking $M=1, N=0$.
In the subsequent section, we use notations $|s_0^{(N)}|$ or $|s_1^{(N)}|$, which means $N$-coupled terms in Eq. (\ref{sollin}).
In this case,  $|s_0^{(N)}|$ is obtained from $|s_0|$ in Eq. (\ref{sollin}) by taking $M=0, N=1$.
As we move away from the soliton, it becomes
$|s_2| >> |s_0^{(M)}| \sim |s_1^{(M)}| $ or $|s_2| <<  |s_0^{(M)}| \sim |s_1^{(M)}| $, and $\j_2$ in Eq. (\ref{newj}) becomes zero.
We can see from Eq. (\ref{sollin}) that the dominating factor of $|s_0^{(M)}|$ (or $|s_1^{(M)}|$) 
is $\exp(-\Im \D)$. Other terms give much small oscillating behavior. 
The dominating factor of  $|s_2|$ is $\exp(\Im (\l^2 \zb-\l z)/2)$.
Thus, the soliton pair moves along the line $z =\a \zb$, where 
\be
\a={\Im (\l^2 /2-\g+pv \b) / \Im (\b p+\l/2) }.
\label{vel}
\ee
The value $\a$ of the soliton for parameters of Figure 1 is  calculated to be -1.92 using Eq. (\ref{vel}),
which is in accordance with Figure 1.

Another interesting feature of Figure 1 ($|\j_1|$) is that the crest of the 
cnoidal background shifts constantly across the dark soliton. The shift is calculated as following.
As we move away from the soliton such that
$|s_2| >> |s_0^{(M)}| \sim |s_1^{(M)}|$,
we find that $\j_1^{c-s} (z,\zb) \rightarrow \j_1 ^c (z,\zb)$ from Eq. (\ref{newj}).
Thus $|\j_1^{c-s}| \rightarrow p \dn \ch$ in this region. 
On the other side of the dark soliton, it becomes $|s_2| <<  |s_0^{(M)}| \sim |s_1^{(M)}|$, and
$\j_1^{c-s} (z,\zb) \rightarrow \j_1 ^c (z,\zb)+i(\l -\l^* )(s_1^{(M)} /s_0^{(M)}+s_0^{(M)*} /s_1^{(M)*} )^{-1}$.
Using that $s_1^{(M)} /s_0^{(M)} = - e^{-i \z} \sn (-i u) \dn (\chi+i u) /  \cn (-i u)$
(Use Eqs. (\ref{sollin}) and (\ref{rel1}) with $N=0$), we find that (For simplicity, we take $u=i u_I$.)
\be
\j_1^{c-s} e^{-i \z} \rightarrow p ~\dn \ch-2 p {\dn u_I \over \cn u_I ~\sn u_I}
\left( {\sn u_I \dn(\ch-u_I) \over \cn u_I}+{\cn u_I \over \sn u_I \dn(\ch-u_I)}\right)^{-1},
\label{shift}
\ee
where we use Eqs. (\ref{lambda}), (\ref{rel2}) and $\sn u_I \equiv \sn (u_I, k)$.
Using the addition theorem, Eq. (\ref{shift}) can be written
\ben
&&p {\dn (\ch-u_I) \dn u_I -k^2 \sn u_I \cn u_I \sn (\ch-u_I) \cn (\ch-u_I) \over 1 -k^2 \sn^2 u_I \sn^2 (\ch-u_I)}
\nonumber \\
&&-2p{\dn u_I \dn (\ch-u_I) \over \sn^2 u_I \dn^2 (\ch-u_I) +\cn^2 u_I} \nonumber \\
&&=p {-\dn (\ch-u_I) \dn u_I -k^2 \sn u_I \cn u_I \sn (\ch-u_I) \cn (\ch-u_I) \over 1 -k^2 \sn^2 u_I \sn^2 (\ch-u_I)}
\nonumber \\
&&= -p \dn (\ch-2 u_I).
\label{shift1}
\een
Thus, $|\j_1^{c-s}| \rightarrow p \dn (\ch-2 u_I)$ in this region, and the shift of crest is $2 u_I$.
For complex values of $u= u_R +i u_I$, we find the following identity which is checked numerically for various values of complex
$u$ and real $\ch, k$.
\ben
 | \dn (\ch,k) - \left( {\dn(-iu,k) \over \cn(-iu,k) \sn(-iu,k)} +{\dn(iu^*,k) \over \cn(iu^*,k) \sn(iu^*,k)} \right) \times \nonumber \\ 
\left( {\sn(-iu,k) \dn(\ch+iu,k) \over \cn(-iu,k)} +
{\cn(iu^* ,k) \over \sn(iu^* ,k) \dn(\ch-i u^* ,k) } \right) ^{-1} |
=\dn(\ch-2 u_I, k).
\label{dnid}
\een
Using Eq. (\ref{dnid}), we find that
$|\j_1^{c-s}| \rightarrow p \dn (\ch-2 u_I)
= p \dn (\ch-2 \Im u)$ still holds.
We see that the parameter $u$
is suitable (compared to the DT parameter $\l$) in describing the characteristics of (soliton + cnoidal) system. 
The shift of the crest in terms of $z$ is $z \rightarrow z+{2 \Im u / p}$
when we take $v=0$ in $\ch=-p(z - v \zb)$. In Figure 1, this shift is $0.97$.

\subsection{Soliton on top of Cnoidal wave}
Recently there arise new interests on optically-induced lattices and the
localization of light in those gratings \cite{efr,fle,des}.
Especially, Ref. \cite{des} describes a stationary configuration of (soliton + cnoidal wave) system
where a soliton moves in parallel with the crest of a cnoidal wave. In this case, the velocity of the soliton is zero
and Eq. (\ref{vel}) requires that $\l^2 /2-\g+pv \b$ should be a real number.
It, in turn, requires that $u$ takes forms $u=i u_I$ or $u=i u_I +K'$. To see this, use Eqs. (\ref{gb}), (\ref{lambda}) and (\ref{rel2}).
Note that \cite{russian}
\be
\sn (a+i K')={1 \over \sn a}, ~\cn (a+i K')=-{i \over k} {\dn a \over \sn a}, ~\dn (a+i K')=-i {\cn a \over \sn a}.
\ee
Here $u_I$ is an real number and $K'= K(\sqrt {1-k^2})$.

Fig. 2 shows one such case. Here solitons run in parallel with the crest of cnoidal wave.
For this figure, we take $M=1.1, N=1$. In this case, there arises interference 
between the $M$ and $N$-coupled terms in $s_0, s_1$
of Eq. (\ref{sollin}). This interference results in an oscillating behavior of solitons as shown in Fig. 2.

\begin{figure}
\centerline{\epsfxsize 5. truein \epsfbox {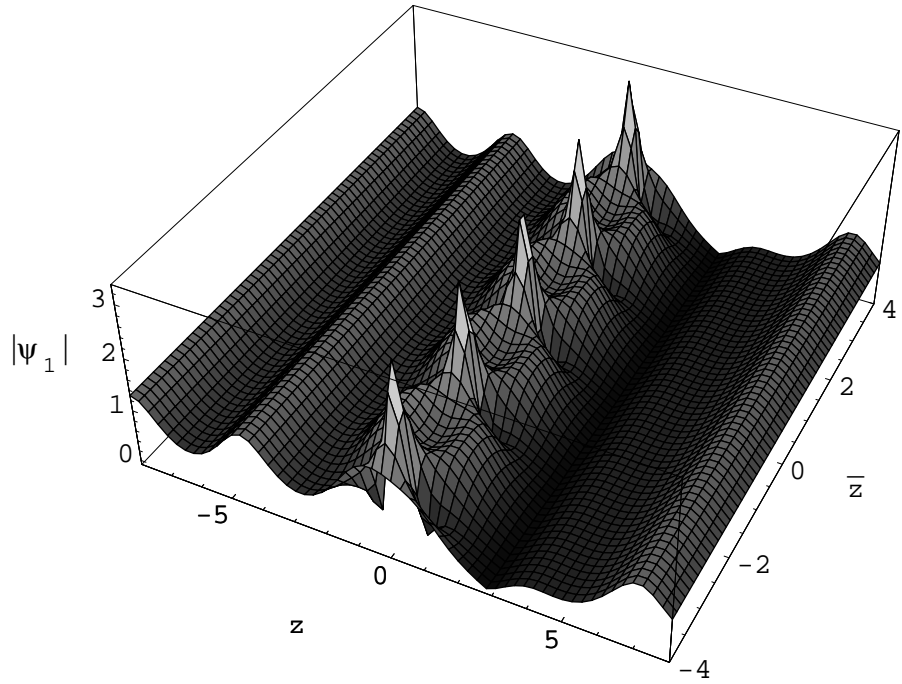}}
\centerline{\epsfxsize 5. truein \epsfbox {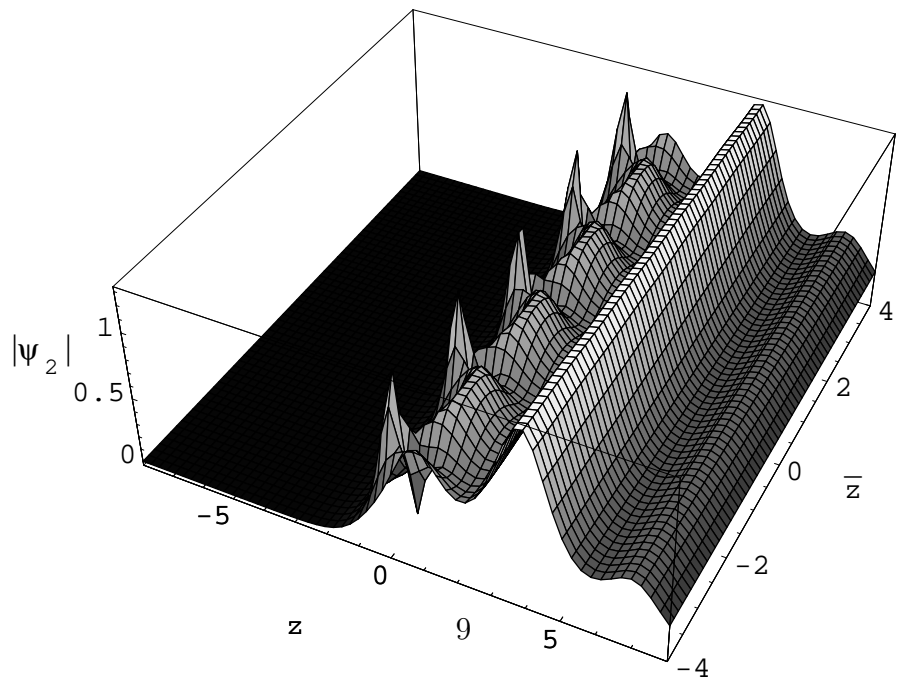}}
\caption{Dark-bright soliton pair moving paralelly along the crest of cnoidal wave: $|\psi_1|$ for dark soliton,
$|\psi_2|$ for bright soliton.
The parameters are $v=0, k=0.9, p=1.3, u=0.63i, M=1.1, N=1, C=0.3.$}
\end{figure}
The oscillating behavior disappears in the case of $N=0$ (or $M=0$), and we
get a stationary configuration of (soliton + cnoidal wave) system. This case corresponds to the system discussed
in \cite{des}. (They use numerical analysis.) Various stationary configurations dealt in \cite{des} can
be obtained from our result in Eqs. (\ref{newj}) and (\ref{sollin}) by taking proper values of parameters.
It includes cases taking  $k \ge 1$, $0 \ge k \ge 1$, as well as taking $u=i u_I$ or $u=i u_I +K'$ with $N=0$ (or $M=0$).
More detailed characteristics of solutions corresponding to this specific choices of parameters
are discussed elsewhere \cite{kiv}.

\subsection{Soliton Fusion on a Cnoidal wave background}
More general solution describing (soliton+cnoidal wave) is given by taking $M \ne 0, N \ne 0$ in Eq. (\ref{sollin}).
Figure 3  describes one of this case. We see that a dark-bright pair breaks up into another dark-bright pair
plus an oscillating soliton.
\begin{figure}
\centerline{\epsfxsize 5. truein \epsfbox {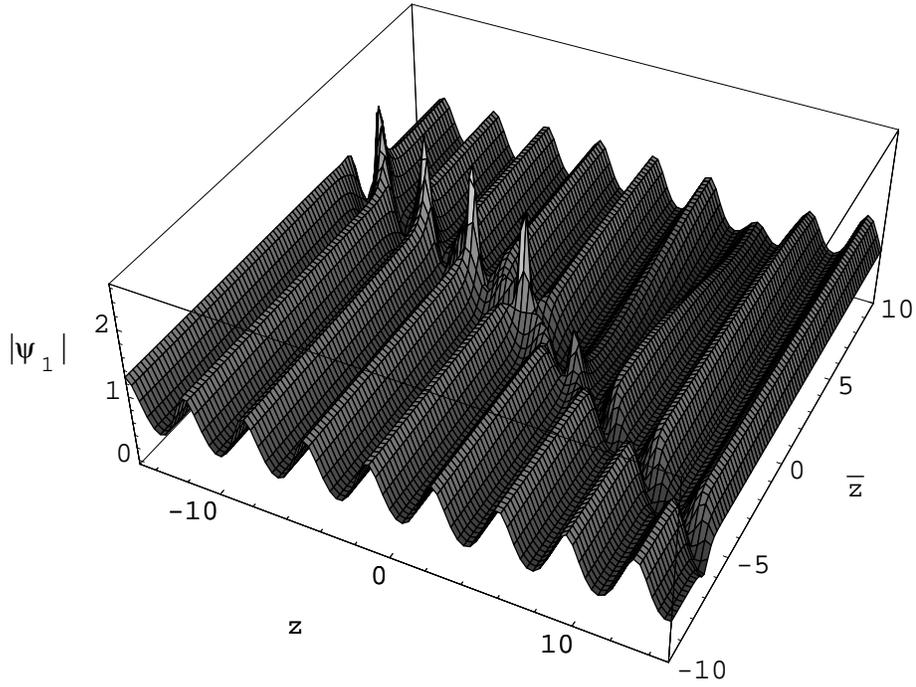}}
\centerline{\epsfxsize 5. truein \epsfbox {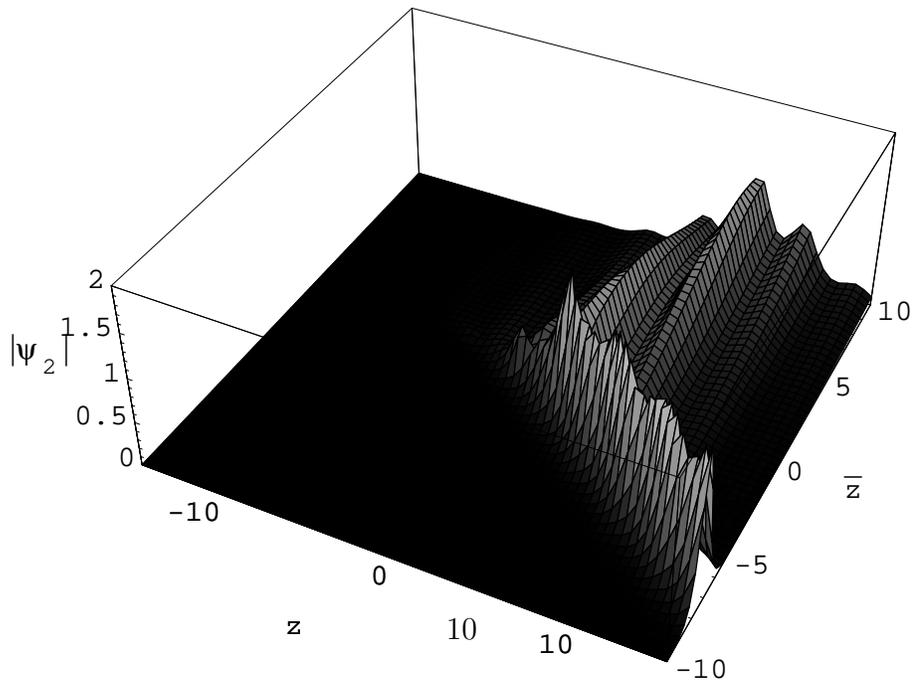}}
\caption{Dark-bright soliton pair breaks up into another dark-bright pair plus an
oscillating soliton: $|\psi_1|$ shows a dark soliton breaking up into a dark plus oscillating solitons.
$|\psi_2|$ shows a bright soliton changing its direction.
The parameters are $v=0, k=0.9, p=1.3, u=-0.38+0.63i, M=1.1, N=1, C=0.3.$}
\end{figure}
This feature was first found in \cite{shin}. Our result is a generalization
of the result in \cite{shin}, such that a dark soliton resides on
a cnoidal wave instead on a constant background. 

There are three soliton lines in Fig. 3. The direction of these lines is calculated similarly as in the subsection 3.1.
The directions of two dark-bright pairs are given by  lines satisfying either $|s_2| \sim |s_0^{(M)}| \sim |s_1^{(M)}|$ 
or $|s_2| \sim |s_0^{(N)}| \sim |s_1^{(N)}|$. The direction
of soliton line for $|s_2| \sim |s_0^{(M)}| \sim |s_1^{(M)}|$ is given in Eq. (\ref{vel}) of subsection 3.1. It was $z=-1.92 \zb$
for parameters of Fig. 1 (or Fig. 3). Contrary to the case of Fig. 1 (where the soliton exists on a full straight line), 
the soliton in Fig. 3 (which exists on a half of the line) exists only in the region
satisfying $|s_2| \sim |s_0^{(M)}| \sim |s_1^{(M)}| >>  |s_0^{(N)}| \sim |s_1^{(N)}|$.
Using the expression for $\j_2$ in Eq. (\ref{newj}), it is easy to see that the soliton disappears
in the region $|s_2| \sim |s_0^{(M)}| \sim |s_1^{(M)}| <<  |s_0^{(N)}| \sim |s_1^{(N)}|$.

Another dark-bright pair moves along a line satisfying $|s_2| \sim |s_0^{(N)}| \sim |s_1^{(N)}|$, or $z =\hat \a \zb$, where 
\be
\hat \a={\Im (\l^2/2+\g-pv \b) \over \Im (-\b p+\l/2)}.
\label{vel1}
\ee
In this case, the dominating factor of $|s_0^{(N)}|$ (or $|s_1^{(N)}|$) in Eq. (\ref{sollin})
is $\exp(\Im \D)$.
The dominating factor of  $|s_2|$ is still $\exp(\Im (\l^2 \zb-\l z)/2)$.
Thus, the soliton pair moves along the line $\Im \D =\Im (\l^2 \zb-\l z)/2)$, which gives Eq. (\ref{vel1}).
For parameters of Fig. 3, $\hat \a=0.44$.

The oscillating soliton in $|\psi_1|$ of Fig. 3 moves along a line $z =\tilde \a \zb$, where 
\be
\tilde \a={\Im (-\g+pv \b) \over \Im (\b p)}.
\label{vel2}
\ee
In this case, the line is described by  $|s_0^{(M)}| (\sim |s_1^{(M)}|) \sim |s_0^{(N)}| (\sim |s_1^{(N)}|)$.
The dominating factor of $|s_0^{(M)}|$ (or $|s_1^{(M)}|$) is $\exp(-\Im \D)$, while
that of  $|s_0^{(N)}|$ (or $|s_1^{(N)}|$) is $\exp(\Im \D)$.
Thus, the oscillating soliton moves along a line $\Im \D =0$, which gives Eq. (\ref{vel2}).
For parameters of Fig. 3, $\tilde \a=-2.89$. 
Note that on a line of the oscillating soliton, it becomes $|s_2| << |s_0| \sim |s_1|$. 
Thus there appears no soliton in $|\j_2|$ along this line, see Eq. (\ref{newj}).
The nature of the oscillating soliton in $\j_1$ is exactly that of the single-component NLSE in \cite{shinpre}.
In the region of $|s_2| >> |s_0| \sim |s_1|$, the oscillating soliton disappears even in $|\j_1|$.

The shift of the crest of the cnoidal wave across the solitons can be similarly calculated as in the subsection 3.1.
Especially in the region (Region I) where $|s_2| >> |s_0^{(M)}| \sim |s_1^{(M)}|$ and $|s_2| >> |s_0^{(N)}| \sim |s_1^{(N)}|$,
we have $|\j_1^{c-s} (z,\zb)| \rightarrow |\j_1 ^c (z,\zb)|= p \dn (\ch)$. In the region (Region II) where
$|s_2| <<  |s_0^{(M)}| \sim |s_1^{(M)}|$ and $|s_0^{(N)}| \sim |s_1^{(N)}| <<  |s_0^{(M)}| \sim |s_1^{(M)}|$, we have
$\j_1^{c-s} (z,\zb) \rightarrow \j_1 ^c (z,\zb)+i(\l -\l^* )(s_1^{(M)} /s_0^{(M)}+s_0^{(M)*} /s_1^{(M)*} )^{-1}$
and $|\j_1^{c-s}| \rightarrow p \dn (\ch-2 u_I)= p \dn (\ch-2 \Im u)$, see subsection 3.1.
In the region (Region III) where
$|s_2| <<  |s_0^{(N)}| \sim |s_1^{(N)}|$ and $|s_0^{(N)}| \sim |s_1^{(N)}| >>  |s_0^{(M)}| \sim |s_1^{(M)}|$, we have
$\j_1^{c-s} (z,\zb) \rightarrow \j_1 ^c (z,\zb)+i(\l -\l^* )(s_1^{(N)} /s_0^{(N)}+s_0^{(N)*} /s_1^{(N)*} )^{-1}$.
In this case, $s_1^{(N)} /s_0^{(N)} = - e^{-i \z} \cn (-i u) / (\dn (\chi-i u) \sn (-i u))$. Similar procedure
used in Eqs. (\ref{shift}) and (\ref{shift1}) gives us $|\j_1^{c-s}| \rightarrow p \dn (\ch+2 u_I)= p \dn (\ch+2 \Im u)$.

Region I and Region II meat at a boundary, along which the dark-bright soliton pair moves with the velocity $\a$ in Eq. (\ref{vel}).
Along the boundary between Region I and III, the soliton pair moves with the velocity $\hat \a$ in Eq. (\ref{vel1}).
Finally, the oscillating soliton moves along the boundary between Region II and III with the velocity $\tilde \a$
in Eq. (\ref{vel2}). Thus the relative shift of the crest of the cnoidal wave across the boundary described by $\tilde \a$ is given by $\ch \rightarrow
\ch -4 \Im u$.
Then the shift of the crest in terms of $z$ is $z \rightarrow z+{4 \Im u / p}$
when we take $v=0$ in $\ch=-p(z - v \zb)$. For parameters of Figure 3, this shift is $1.94$.

\section{Solution of defocusing medium}
Figure 4 shows a dark soliton on a cnoidal background in a defocusing medium. It is obtained using
Eqs. (\ref{j121}), (\ref{newj}), (\ref{sollin}) and taking $\s =-1$,
$v=0, k=0.9i, p=1.3i, u=-0.3-0.98i, M=1, N=0, C=0.1$. Note that in the case of defocusing medium ($\s=-1$),
we take imaginary $k, p$.
Another specific feature of the defocusing medium is that
it is not permitted to take both $M$ and $N$ to be simultaneously nonzero. This fact is explained as following.
\begin{figure}
\leftline{\epsfxsize 5. truein \epsfbox {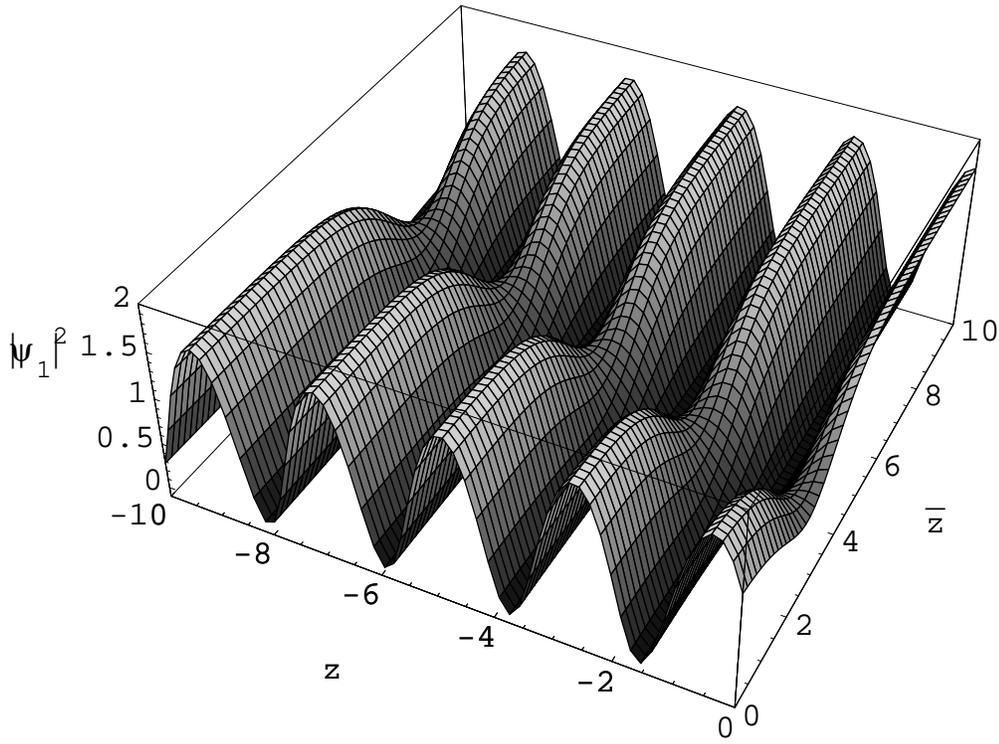}}
\rightline{\epsfxsize 5. truein \epsfbox {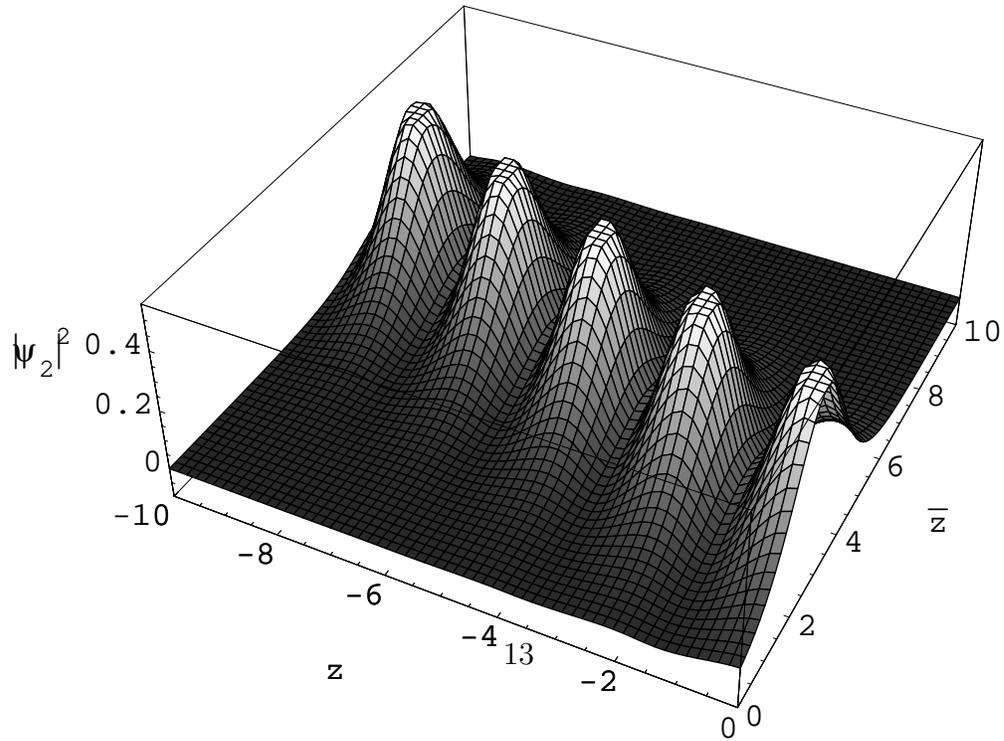}}
\caption{Dark-bright soliton pair moving in a defocusing medium: $|\psi_1|^2$, a dark soliton and 
$|\psi_2|^2$, a bright soliton.
The parameters are $v=0, k=0.9i, p=1.3i, u=-0.3-0.98i, M=1, N=0, C=0.1$}
\end{figure} 
The denominator part of Eq. (\ref{sollin}) is $|s_0|^2-|s_1|^2-|s_2|^2$, which is required
to be positive definite or negative definite on all $(z, \zb)$ plane.
Otherwise, we will get an unphysical singular solution. But
there always exists a region in the $(z, \zb)$ plane where $|s_2| >> |s_0|$, which means the denominator
must be negative definite. Thus we need $|s_0|^2-|s_1|^2$ should be negative definite.
In the region $\Im \D \sim -\infty$ where $|s_0| \sim |s_0^{(M)}|, |s_1| \sim |s_1^{(M)}|$,
the negative definiteness requires $1<|s_1^{(M)} /s_0^{(M)} | = |\sn (-i u) \dn (\chi+i u) /  \cn (-i u)|$.
On the other hand, in the region $\Im \D \sim \infty$ where $|s_0| \sim |s_0^{(N)}|, |s_1| \sim |s_1^{(N)}|$,
we need $1<|s_1^{(N)} /s_0^{(N)}| = |\cn (-i u) / (\dn (\chi-i u) \sn (-i u))|$. As $|\dn (\chi \pm i u)|$
is an oscillating function of $\chi$,  the negative definiteness from the two regions results
in a contradiction on the magnitude of $|\sn (-i u) /  \cn (-i u)|$. Thus we need one of $M, N$ should be zero.
Thus there does not exist the phenomenon of soliton fusion in the case of defocusing medium.

When we take $N=0$ as in Fig. 4, we need $1<|\sn (-i u) \dn (\chi+i u) / \cn (-i u))|$.
In the appendix II, we prove that 
\be
(1+k^2)^{(1/4)} \left|{\sn (-i u_R +m \tilde K' ,ik) \over \cn (-i u_R +m \tilde K' ,ik)}\right|
= |\dn(\chi+i u_R -m \tilde K' ,ik)|/(1+k^2)^{(1/4)} =1,
\label{ide}
\ee
where 
\be
\tilde K'  = {K'(1/ \sqrt{1+k^2}) \over 2 \sqrt{1+k^2}},
\label{tkp}
\ee
$m$ is an odd number ($m=2n+1$), $\chi=-ip (z-v \zb)$ and $k, p, v, u_R$ are real numbers.
(Here, we use the fact that $k \rightarrow i k, p\rightarrow i p$ in the defocusing medium.)
We also found that 
\be
(1+k^2)^{(1/4)} \left|{\sn (-i u_R +u_I,ik) \over \cn (-i u_R+u_I,ik)} \right|>1, ~|\dn(\chi+i u_R-u_I,ik)|/(1+k^2)^{(1/4)} >1,
\label{ineq}
\ee
when
\be
 (4 n+1) \tilde K' < u_I < (4n+3) \tilde K'
\label{ineq1}
\ee
where $n$ is an arbitrary integer. On the outsides of interval in Eq. (\ref{ineq1}), Eq. (\ref{ineq}) reverses its inequality sign.
Thus we can see that the negative definiteness constrains the possible value of $u_I$ to the intervals given in Eq. (\ref{ineq1}).
Similarly we must take the value $u_I$ in the outsides of intervals in Eq. (\ref{ineq}) when
we take $M=0$. For the parameters of Fig. 4, $\tilde K'=0.67$.

The dark-bright pair soliton moves along a line, which satisfies $|s_2| \sim |s_0^{(M)} | \sim |s_1^{(M)} | $ ($N=0$ case).
Thus we can use Eq. (\ref{vel}) to describe the direction of the soliton line. In the case of defocusing medium, we must take
imaginary $p, k$ values. For parameters of Fig. 4, $\a=-2.48$. The soliton line shifts along the $\zb$-axis (or $z$-axis)
according to the relative magnitude of $M$ and $C$. This shift is rather large in Fig. 4 compared to the other figures,
which is due to large $M/C =10$ value. 

The shift of the crest for the case of defocusing medium is similarly calculated as in subsection 3.1.
At the region $|s_2| >> |s_0^{(M)} | $, $\j_1^{c-s} (-pi(z-v \zb) )=\j_1^{c} (-pi(z-v \zb) )$.
To calculate  $\j_1^{c-s} $ in the region the $|s_2| << |s_0^{(M)} | $, we use the following formula
\ben
&&| \dn (i \ch,ik) - \left( {\dn(-iu,ik) \over \cn(-iu,ik) \sn(-iu,ik)} -{\dn(iu^*,-ik) \over \cn(iu^*,-ik) \sn(iu^*,-ik)} \right) \times \nonumber \\ 
&&\left( {\sn(-iu,ik) \dn(i\ch+iu,ik) \over \cn(-iu,ik)} -
{\cn(iu^* ,-ik) \over \sn(iu^* ,-ik) \dn(-i\ch-i u^* ,-ik) } \right) ^{-1} | \nonumber \\
&&=|\dn(i\ch+2i \Re u, ik)|,
\label{dnid1}
\een
where $u$ is complex and $\ch, k$ are real.
Then using Eqs. (\ref{newj}) and (\ref{dnid1}), we find that
$|\j_1^{c-s} (-pi(z-v \zb) )|=|\j_1^{c} (-pi(z-v \zb) +2i \Re u)|$ for real $p$.
The shift of the crest in terms of $z$ is $z \rightarrow z-{2 \Re u / p}$
for the case of $v=0$. In Figure 4, this shift is $0.46$.

\section{Conclusions}
In this paper, we have introduced (soliton+cnoidal wave) solutions of
the CNLS equation. They were obtained using  the DT and introducing a solution of the associated linear problem
on a cnoidal background. We calculate the moving direction of a soliton on a cnoidal background and
the shift of the crest of a cnoidal wave. We also found a solution in the self-focusing case where a 
dark-bright pair 
breakup into another dark-bright pair and an oscillating soliton. 
These type of solutions, which can be easily applicable to the analysis of
physically interesting processes, seem rather rare in the literature of physics. These solutions
can be used, for example, in describing the localized states in optically induced refractive index gratings.
 
The stability analysis of these solutions is remained for future study. In fact, there already appears some numerical
studies on this subject \cite{des}. There was shown that the (soliton+ cnoidal wave) system is unstable or weakly stable
for the focusing case, while it is stable for the defocusing case.

\vglue .2in
\centerline{\bf ACKNOWLEDGMENT}
\vglue .2in
I am grateful to Yuri S. Kivshar for showing me the paper \cite{des} prior to publication.
This work was supported by Korea Research Foundation Grant (KRF-2003-070-C00011).

\vglue .2in

\section{Appendix I: Proof of Sym's solution }

In this appendix, we show that $s_0, s_1, s_2$ in Eq. (\ref{sollin}) indeed satisfy the linear equation (\ref{nlsle}).
Consider the following equation, which is obtained from the $\pp$-part of Eq. (\ref{nlsle});
\be
\pp s_0 +\j_1 ^{c} s_1 +i \l s_0 /2=0.
\label{fst}
\ee
By inserting $s_0, s_1$ (For simplicity, we consider a case of $M=1, N=0$ in Eq. (\ref{sollin}).) into Eq. (\ref{fst}) and
taking $\j_1 ^{c}$ as in Eq. (\ref{j121}), we get
\be
-ip \b -{p \over 2K} {\q_0 '({\ch+iu \over 2K}) \over \q_0 ({\ch+iu \over 2K})}
+{p \over 2K} {\q_0 '({\ch \over 2K}) \over \q_0 ({\ch \over 2K})} 
-p ~\dn(\ch,k) {\q_1 ({-iu \over 2K}) \over \q_2 ({-iu \over 2K})}  {\q_3 ({\ch+iu \over 2K}) \over \q_0 ({\ch+iu \over 2K})}
 +p {{\rm dn}(u,k') {\rm cn}(u,k')
\over {\rm sn}(u,k')}.
\label{sst}
\ee
Inserting $\b$ in Eq. (\ref{gb}) into Eq. (\ref{sst}) and using
following identities \cite{jms,russian},
\be
\int_0 ^u \dn^2 u ~du ={1\over 2K} {\q_0 '({u \over 2K}) \over \q_0 ({u \over 2K})} +{E \over K} u,
\ee
\be
{\rm sn} u={1 \over \sqrt k} {\q_1({u \over 2K}) \over \q_0 ({u \over 2K})},~
{\rm cn} u={\sqrt {k'} \over \sqrt k} {\q_2({u \over 2K}) \over \q_0 ({u \over 2K})},~
{\rm dn} u=\sqrt {k'} ~{\q_3({u \over 2K}) \over \q_0 ({u \over 2K})},
\label{rel1}
\ee
\be
\sn(iu,k') =i {\sn(u,k) \over \cn(u,k)},~\cn(iu,k') = {1 \over \cn(u,k)},~\dn(iu,k') = {\dn(u,k) \over \cn(u,k)},
\label{rel2}
\ee
we get
\be
-p \int^{\ch+iu} _0 [\dn^2 v -\dn ^2 (v-iu) ] dv +{\sn(-iu,k) \over \cn(-iu,k)}
[\dn(-iu,k)- \dn (\ch,k) ~\dn(\ch+iu,k)].
\label{tst}
\ee
Finally, using the following identity (It is a result of the addition theorem of Jacobi's elliptic functions.)
\ben
&&\dn^2 (a-b) -\dn ^2 a = k^2 {\sn b \over \cn b} [\cn a ~\sn a ~\dn (a-b) +
\dn a ~\cn (a-b) ~\sn (a-b)] \nonumber \\
&=&-{\sn b \over \cn b} {d \over d a}[\dn a ~\dn (a-b)],
\een
we can see that Eq. (\ref{tst}) becomes zero. 
Other $\pp$-part of the linear equation (\ref{nlsle}) (including the $M=0,N=1$ case of $s_0, s_1$ in Eq. (\ref{sollin}))
are similarly proved.

A $\pb$-part of Eq. (\ref{nlsle}) is ($\s=1$ case)
\be
\pb s_0 +i |\j_1 ^{c}|^2 s_0 -i  \pp \j_1 ^{c} ~s_1 -\l \j_1 ^{c} s_1- i \l^2 s_0 /2=0.
\label{bst}
\ee
By inserting $s_0, s_1$ (For simplicity, we take $M=1, N=0$ in Eq. (\ref{sollin}).) into the left part of Eq. (\ref{bst}) and
taking $\j_1 ^{c}$ as in Eq. (\ref{j121}), we get
\ben
{i \over 2} [-p^2 (2-k^2) +{v^2 \over 4}]-i \g -v ({\pp s_0 \over s_0} +i {v \over 4})
+i p^2 \dn^2 \ch-{i \over 2}  \l^2 \nonumber \\
+~p ~[~i \pp \dn \ch +(\l +{v\over 2} ) \dn \ch]  ~{\q_1 ({-iu \over 2K}) 
\over \q_2 ({-iu \over 2K})}  {\q_3 ({\ch+iu \over 2K}) \over \q_0 ({\ch+iu \over 2K})}.
\label{bst1}
\een
Using Eq. (\ref{fst}) and the identity (\ref{rel1}), Eq. (\ref{bst1}) becomes
\ben
&&-{i \over 2} p^2 (2-k^2) +{i \over 8} v^2 +i {p^2 \over 2} [\dn^2(u ,k')+{\cn^2(u, k') \over \sn^2 (u ,k')}]
+{i \over 2} \l v-{i \over 4} v^2-{i \over 2} \l^2 \nonumber \\
&& +i p^2 \dn^2 \ch -({p \over 2} v \dn \ch-i p^2 k^2 \sn \ch \cn \ch-\l p \dn \ch)
{\sn(-iu, k) \over \cn(-iu, k)} \dn(\ch+iu, k).
\label{bst2}
\een
With the help of Eqs. (\ref{rel2}) and (\ref{lambda}), repeated applications of addition theorem
on Eq. (\ref{bst2}) gives zero, which proves Eq. (\ref{bst}).
Other $\pb$-part of the linear equation (\ref{nlsle}) (including the $M=0,N=1$ case of $s_0, s_1$ in Eq. (\ref{sollin}))
are similarly proved.

\section{Appendix II: Proof of Eq. (17) }
We first start with 
\be
(1+k^2)^{(1/4)} {\sn (-i u, ik) \over \cn (-i u, ik)}= i \sn(-\sqrt{1+k^2} u, \tilde k ) / (1+k^2)^{(1/4) },
\label{eq1}
\ee
where $\tilde k ={1 / \sqrt{1+k^2}}$ with $k$ real and $u$ is a complex value.
By inserting $u=u_R +i \tilde K'$ in Eq. (\ref{tkp}) and using the addition theorem, Eq. (\ref{eq1}) becomes
\be
{ \sn( -\sqrt{1+k^2} u_R, \tilde k  ) (1+\tilde k )-i \cn(-\sqrt{1+k^2} u_R, \tilde k ) 
\dn(-\sqrt{1+k^2} u_R, \tilde k ) 
 \over 1+\tilde k \sn(-\sqrt{1+k^2} u_R, \tilde k ) },
\label{eq2}
\ee
where we use 
\ben
&&\sn (-i \sqrt{1+k^2} \tilde K' , \tilde k ) =-i (1+k^2)^{(1/4)},~\dn (-i \sqrt{1+k^2} \tilde K' , \tilde k ) =\sqrt{1+ \tilde k }, 
\nonumber \\
&&\cn (-i \sqrt{1+k^2} \tilde K' , \tilde k  ) = (1+k^2)^{(1/4)} \sqrt{1+ \tilde k }.
\een 
It is now easy to see that the absolute value of Eq. (\ref{eq2}) is 1.

In the case of $\dn$ formula,
\be
\dn (-i u, ik) / (1+k^2)^{(1/4)} = {\cn(-\sqrt{1+k^2} u, \tilde k )  \over \dn(-\sqrt{1+k^2} u, \tilde k ) (1+k^2)^{(1/4)}}.
\label{eq3}
\ee
Applying a similar procedure used in obtaining Eq. (\ref{eq2}) to Eq. (\ref{eq3})  gives
\be
{\cn(-\sqrt{1+k^2} u_R, \tilde k ) +i  \sn(-\sqrt{1+k^2} u_R, \tilde k )   \dn(-\sqrt{1+k^2} u_R, \tilde k ) 
\over  \dn(-\sqrt{1+k^2} u_R, \tilde k ) 
-i  \tilde k \cn(-\sqrt{1+k^2} u_R, \tilde k )  \sn(-\sqrt{1+k^2} u_R, \tilde k ) },
\ee
whose absolute value is 1. It is easy to expand above results to the case of $u=u_R-i \tilde K'$.
Using that $\sn (-i u, ik) / \cn (-i u, ik), \dn (-i u, ik)$ are periodic functions in $u_I$
with the period $2 \tilde K'$, we can obtain the result in Eq. (\ref{ide}) in subsection 4.

To obtain a result shown in Eq. (\ref{ineq}), we should go through similar procedures in the above paragraphs with much complex expressions
in this case. Instead of going through a tedious calculation, we use MAPLE to check numerically the validity of Eq. (\ref{ineq}).

\end{document}